\documentclass[english,groupedaddress, superscriptaddress,aps,pre,10pt, nofootinbib, notitlepage,longbibliography]{revtex4-2}
\usepackage{mathtools}
\usepackage{amsmath}
\usepackage{amssymb}
\usepackage{bigints}
\usepackage{graphicx}
\usepackage{gensymb}
\usepackage{float}
\usepackage[normalem]{ulem}
\usepackage{comment}
\makeatletter

\renewcommand*{\p@subsection}{}

\renewcommand*{\p@subsubsection}{}  
\makeatother
\usepackage{geometry}
\usepackage{bigints}
\usepackage{graphicx}
\usepackage{gensymb}
\usepackage{mathtools}
\usepackage{amsmath}
\usepackage{amssymb}
\usepackage[T1]{fontenc}
\usepackage[utf8]{inputenc}

\DeclarePairedDelimiter\abs{\lvert}{\rvert}%
\DeclarePairedDelimiter\norm{\lVert}{\rVert}%

\usepackage{color}
\definecolor{blue}{rgb}{0,0,1}
\definecolor{black}{rgb}{0,0,0}

\definecolor{dgreen}{rgb}{0,0.5,0}

\definecolor{dred}{rgb}{0.5,0,0}

\definecolor{dyellow}{rgb}{0.75,0.75,0}

\makeatletter
\let\oldabs\abs
\def\abs{\@ifstar{\oldabs}{\oldabs*}}
\let\oldnorm\norm
\def\norm{\@ifstar{\oldnorm}{\oldnorm*}}
\makeatother

\makeatletter
\usepackage{enumerate}%
\usepackage[caption=false]{subfig}

\begin{document}
\title{Analysing four major air pollutants' statistics in Europe - A superstatistical approach with q-exponential distribution}
\title{Systematic statistical analysis of air pollution}
\title{Spatial heterogeneity of air pollution statistics}

\author{Hankun He}
\affiliation{School of Mathematical Sciences, Queen Mary University of London, London E1 4NS, United Kingdom}

\author{Benjamin Schäfer}
\thanks{contributed equally}
\affiliation{School of Mathematical Sciences, Queen Mary University of London, London E1 4NS, United Kingdom}
\affiliation{Institute for Automation and Applied Informatics, Karlsruhe Institute of Technology, 76344 Eggenstein-Leopoldshafen, Germany}
\affiliation{Faculty of Science and Technology, Norwegian University of Life Sciences, 1432 Ås, Norway}

\author{Christian Beck}
\thanks{contributed equally}
\affiliation{School of Mathematical Sciences, Queen Mary University of London, London E1 4NS, United Kingdom}
\affiliation{The Alan Turing Institute, London NW1 2DB, United Kingdom}

\begin{abstract}
Air pollution is one of the leading causes of death globally, and continues to have a detrimental effect on our health. In light of these impacts, an extensive range of statistical modelling approaches has been devised in order to better understand air pollution statistics. However, the time-varying statistics of different types of air pollutants are far from being fully understood. The observed probability density functions (PDFs) of concentrations depend very much on the spatial location and on the pollutant substance. 
In this paper, we analyse a large variety of data from 3544 different European monitoring sites and show that the PDFs of nitric oxide ($NO$), nitrogen dioxide ($NO2$) and particulate matter ($PM10$ and $PM2.5$) concentrations generically exhibit heavy tails and are asymptotically well approximated by $q$-exponential distributions with a given width parameter $\lambda$. We observe that the power-law parameter $q$ and the width parameter $\lambda$ vary widely for the different spatial locations. For each substance, we find different patterns of parameter clouds in the $(q, \lambda)$ plane. These depend on the type of pollutants and on the environmental characteristics (urban/suburban/rural/traffic/industrial/background). This means the effective statistical physics description of air pollution exhibits a strong degree of spatial heterogeneity.
\end{abstract}

\maketitle
\makeatother

\section*{Introduction}
Air pollution is among the highest contributors to premature death and disease worldwide, causing a significant number of deaths from stroke, lung cancer and heart diseases~\cite{WHOPollution,Shah}. Besides human health, air pollution affects vegetation, natural ecosystems, climate change, the built environment and subsequently the economy~\cite{Ortiz}. In Europe, air pollution is the single largest environmental health risk~\cite{health2019state,world2018global,gbd2017global} and its long-term effects are very serious. Although emissions and ambient concentrations have fallen steadily in Europe over the past few decades, as stated in \cite{union2008directive,union2016directive}, many European countries still exceed European Union's (EU) standards~\cite{union2008directive} and World Health Organization's (WHO) guidelines, see \cite{world2006air} for the levels of air pollutants in 2018. Two key air pollutants, namely, particulate matter ($PM$) and nitrogen oxides ($NOx$), pose a considerable threat to the health of citizens. About $55000$ and $417000$ premature deaths in 41 European countries in 2018 were attributed to $NO2$ and $PM2.5$, respectively~\cite{Ortiz}. Pollutants such as ozone ($O3$), sulfur dioxide ($SO2$) and carbon monoxide ($CO$) negatively affect human health as well. Particularly, ground level $O3$ has been estimated to have caused $20600$ premature deaths in Europe in 2018; this yearly number has risen by $20\%$ since 2009~\cite{Ortiz}. In this paper we focus on two of the most dangerous pollutants, $NOx$ and $PM$, but the methodologies presented in our paper can be similarly applied to other substances.\par  

The impact of air pollution on health does not only depend on the pollutant type but also on the type of surrounding environment, i.e. people living next to traffic-heavy roads or industries face higher exposure to air pollution. The EU~\cite{european2011directive} uses environmental surrounding types to classify air quality monitoring sites into traffic, industrial, background, urban, suburban and rural, based on predominant emission sources and building density.
From a policy perspective, this allows for evaluating the effectiveness of measures targeting specific emissions sectors and assessing the impact of those associated pollutants which dominate the area surrounding a given monitoring station, such as for example traffic or industry (or their absence). Despite progress made by EU and UK policies~\cite{union2008directive,union2016directive} addressing all sectors to reduce emissions and protect citizens from pollutants, meeting the emission reduction commitments by 2030 remains a challenge~\cite{Cleanair}. Similar challenges to satisfy given policies are also existing in other parts of the world.
 
Having a thorough understanding of the time-varying statistics, i.e. of the entire probability density function (PDF) of air pollution, is crucial for policymakers involved in defining thresholds or reducing overall exposure to air pollution. It is also crucial for
the construction of suitable statistical physics models. PDFs such as gamma, log-normal and Weibull distributions~\cite{Hsin} have been widely used for fitting air pollutant concentration data. However, these distributions decay approximately like exponential functions at large values, while earlier investigations have found heavy tails in air pollution statistics \cite{williams_superstatistical_2020}, which are not well-described by the above distributions. Some recent studies~\cite{schafer2020covid,BALDASANO2020140353,environmental2020effect} have explored the COVID-19 lockdown effects on air quality (in Europe and in megacities such as Delhi), focusing on comparing the PDFs or given moments of the PDFs before and during the lockdown. Superstatistical methods, originating from turbulence modelling~\cite{PhysRevLett.98.064502} and applied to many fields \cite{metzler2020superstatistics, beck2020nonextensive}, offer a powerful effective approach to describe the dynamics of air pollution assuming the existence of well-separated time scales~\cite{williams_superstatistical_2020}. Air pollutants such as $NOx$ have been dealt with success in the superstatistical approach, taking into account nonequilibrium situations with fluctuating variance parameters \cite{williams_superstatistical_2020}. However, this approach has been verified for limited data sets only, chosen from the UK (London), and also
only for a limited set of pollutants, mainly $NO$ and $NO2$.
On a European scale, and for much larger data sets, it remains unclear whether heavy tails are generically observed and whether an effective  superstatistical description is applicable. This is the topic of this paper.

The above consideration leads us to a problem that is of general interest for statistical physics approaches to environmental science. Can we apply standard methods from nonequilibrium statistical physics, such as the above-mentioned superstatistical methods, to environmentally relevant time series of pollution concentrations, and if yes, how can we extract the corresponding superstatistical parameters from the time series? And how spatially heterogeneous are the observed results? What are the values of the relevant superstatistical parameters for different air pollutants and different geographical environments? Furthermore, what are typical distributions of observed PDF fitting parameters for the large number of sites distributed across the European continent?
These important types of questions relating to
large ensembles of different measuring stations will be dealt with in the following. 

The paper is organized as follows. First, we introduce our large data set involving 3544 measuring sites. 
Next, we investigate the relation between mean and standard deviation for the observed PDFs to clarify if the PDFs can be approximated by simple exponential distributions or if more complicated functions are needed. We then systematically investigate the PDFs of all sites, in particular the tail behaviour, and show that the tails are generally much better described by $q$-exponential functions with a given width parameter $\lambda$ than by functions such as exponential and log-normal, meaning there is generically power-law decay. 
Subsequently, we use the maximum likelihood estimation method (MLE) to extract the $q$ and $\lambda$ parameters for the best-fitting $q$-exponential distribution, and present plots of scattered points in the $(q, \lambda)$ plane which exhibit interesting patterns for our large number of spatial locations investigated. Our main result is that air pollution statistics is extremely heterogeneous, with the local variations of best-fitting parameters spanning many orders of magnitudes. Our investigation is the first one that investigated this in a systematic way for very large ensembles of different measuring stations. We show that there is a complex pattern structure in the 2-dimensional ($q$, $\lambda$) parameter space that depends both on the pollutant type as well as on the classification type of the local surroundings.


\section*{Results}

\subsection*{The data set considered}
In this paper, we aim to conduct a large-scale statistical analysis of air pollutants, typically $NOx$ and $PM$, on a European scale. Technically, we access our air quality monitoring data from a large number of locations in Europe through the interface "Saqgetr", which is an R package available on the Comprehensive R Archive Network (CRAN)~\cite{saqgetr}. The vast majority of the accessible data are openly available from the European Commission’s Airbase and air quality e-reporting (AQER) repositories~\cite{EU_Parliament_Council,EU_Commission}. To utilise the data efficiently, they have been processed into a harmonised form with consistent and careful treatment of the observations and metadata by Stuart K. Grange~\cite{saqgetr}. Concentration level readings and site environment types are the two key quantities we investigate. We import 9698 locations' data throughout Europe within the time span of January 2017 to December 2021, recorded at 1-hour intervals. To minimise the influences of seasonal fluctuations in time series, we eliminate sites whose data are too short, typically less than one year. We also exclude sites where a high percentage of measurements falls below the detection limit, since sites with clean air are not our primary analysis goal. Furthermore, we filter out sites whose data are corrupted, the used code and further details are described in the Methods section. We arrive at 3544 sites with data that meet our criteria before we proceed with our statistical analysis. Each data set contains at least $8760~(24\times365)$ data points, up to about $43800~(5\times24\times365)$ if the full 5 year period is available.\par  

\begin{figure}[h]
    \includegraphics[width=13.5cm]{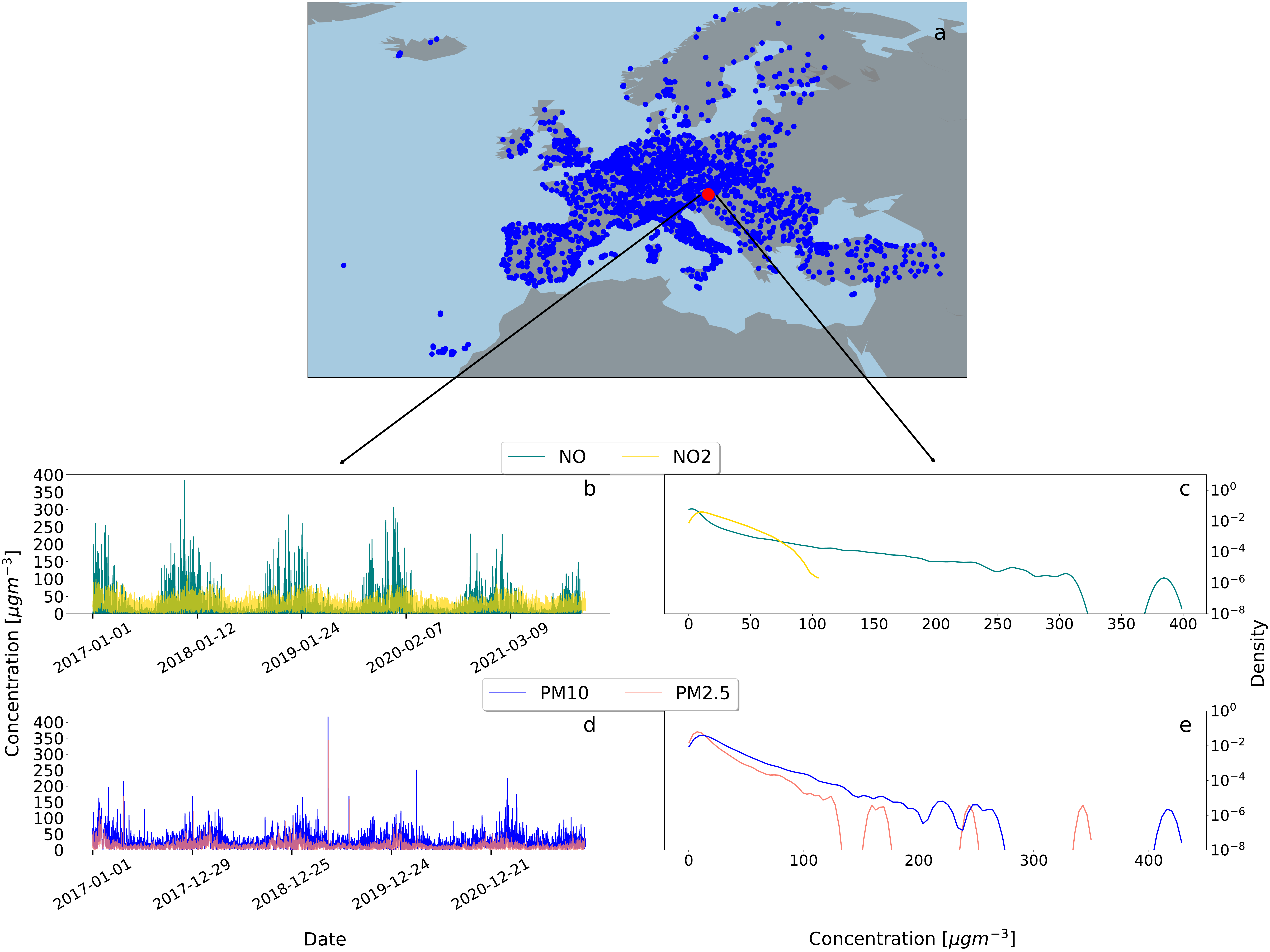} 
    \caption{
    (a) Illustration of the available data sites on a map of Europe, with the red circle labeling our example site: Bahnhofstrasse, Weiz, Austria. Measured time series are shown in (b) and (d), the corresponding probability densities in (c) and (e). All pollutants display clear seasonality in their trajectories.
    \label{fig:map}}
\end{figure}

To provide a general overview of our analysed data, we show all the data sites' locations in Fig.~\ref{fig:map}(a) as well as an example time series of a selected site: Bahnhofstrasse, Weiz in Austria for illustration purposes. 
Measured concentration time series and histograms are shown in Fig.1(b)-(e),
for $NO$, $NO2$, $PM10$, and $PM2.5$. $NOx$ and $PM$ show seasonal cycles, i.e. during winter higher pollutant concentrations are more common. We also observe that for this example site the probability density of $NO$ decays at a slower rate to zero than those of the other three pollutants. Apparently, typical distributions exhibit some heavy tails, which we will analyze in much more detail in the following.


Instead of considering the full distribution of each site's pollutants, in the following we will concentrate onto the tails. One reason for doing so is that we are particularly interested in the statistics of high pollution states, which are most damaging and described by the tails of the distribution. 

With data sets from 3544 air pollution monitoring sites we require an efficient and context-based automated approach of analysing the sites, details are described in the Methods section. Refs \cite{union2008directive,european2011directive} give more details on macro- and micro-scale sampling. In brief, stations are divided into three categories: traffic, industrial, and background based on predominant emission sources; the surrounding areas are classified as urban, suburban, or rural based on the density/distribution of buildings. Station types are combined with area types to provide an overall station classification, and we analyse our data conditioned on this station classification. We use the following definitions: "urban traffic": a site located in close proximity to a busy road in a continuously built-up urban area; "suburban/rural industrial": a site whose pollution level is influenced predominantly by emissions from an industrial area or an industrial source in largely built-up or remote areas; "rural background": a site whose pollution level is influenced by the combined contribution from all sources upwind of the station and not in built-up areas. Eventually, seven environmental area types "urban traffic", "suburban/rural traffic", "urban background", "suburban background", "rural background", "urban industrial" and "suburban/rural industrial" are used in our statistical analysis.

\subsection*{Checking Exponentiality}
Let us first consider the simplest possible hypothesis,
namely that pollution concentrations follow an exponential 
distribution. In this case the PDF is given by
\begin{equation}
  f_\lambda (x) = 
    \begin{cases}
    \lambda e ^ {-\lambda x}  & x\ge0\\
    0 & x<0
    \end{cases}.
    \label{eq:exp}
\end{equation}
For exponential distributions one has the general fact that
\begin{equation}
\mbox{mean}=\mbox{standard deviation} = \frac{1}{\lambda}.
\label{equation 2}
\end{equation}
Thus, for each of our 3455 measuring stations we can easily
test the hypothesis of an exponential distribution by
plotting mean versus standard deviation for the measured data. If pollutants
were to follow an exponential distribution, we would
expect a clustering along the diagonal in such a plot.
Stations with larger $\lambda$ (smaller mean and variance) would correspond to cleaner air, and are expected to be found
closer to the origin (0,0) as compared to highly polluted locations. Our results
are shown in Fig.~2.
\begin{figure}[h]
    \centering
    \includegraphics[width=11cm]{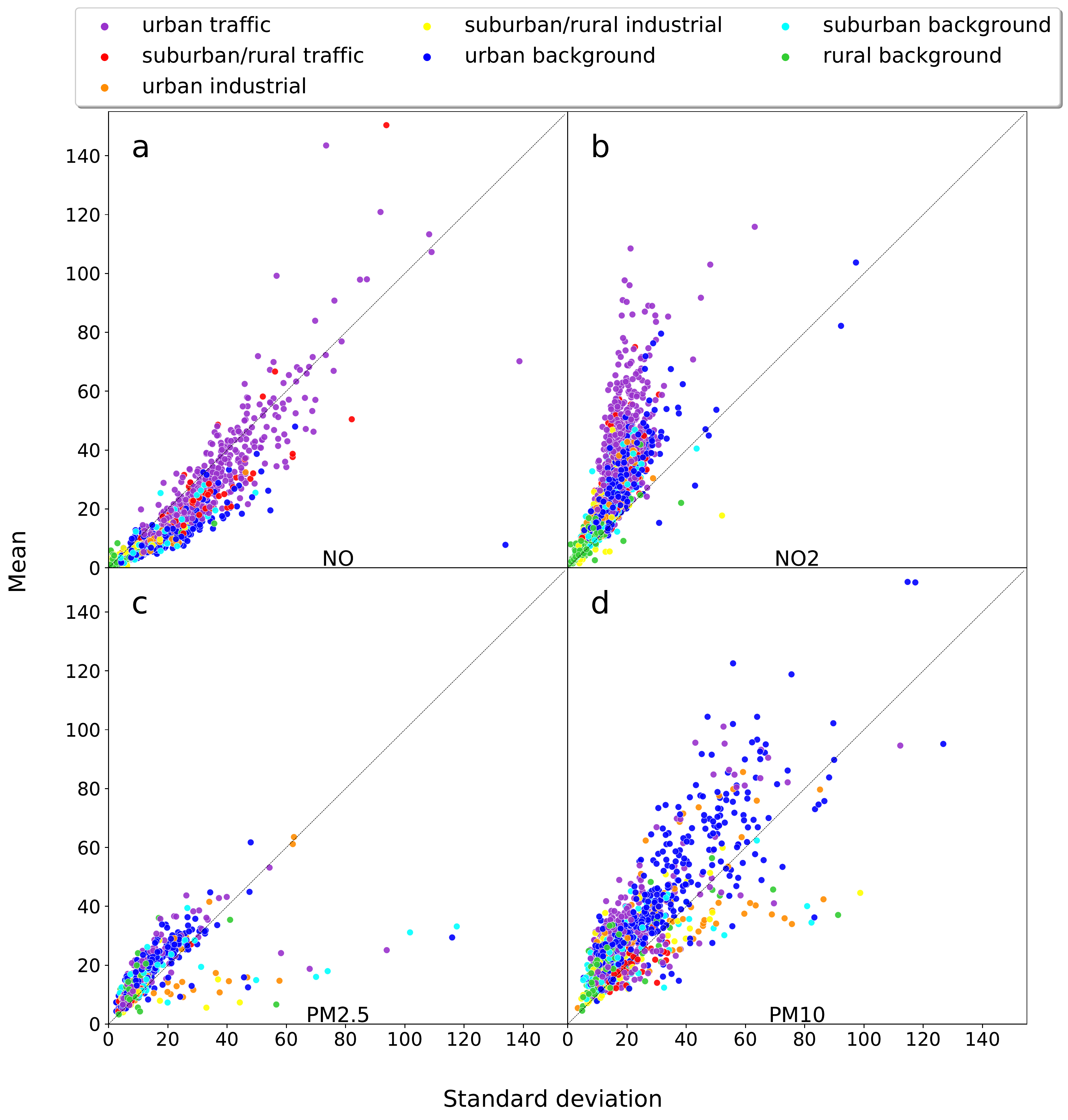}
    \caption{
    For each of the pollutants and measuring stations, we plot mean versus standard deviation. The area type surrounding the measuring station is color-coded.
    Data do not follow an exponential distribution, as evidenced by the fact that the majority of dots do not fall onto the diagonal lines. 
    Different patterns are observed for the four different substances $NO$, $NO2$, $PM2.5$, $PM10$. Green colors (rural stations with clean air) cluster near the origin. However, clustering of the same color patches is observed to be stronger for $NOx$ as compared to $PM$.
    \label{fig:mean_vs_std}}
\end{figure}
The majority of the points are clustered above (b-d) or below (a) the diagonal lines, indicating deviations from an exponential distribution. These deviation patterns are different for each of the four substances. 

Apparently, the PDFs of $NO$ and $NO2$ are very different,
as the points scatter mainly below (NO) and mainly above (NO2) the diagonal. The scattering plots for $PM2.5$ and $PM10$
are more centered around the diagonal, but there are some
unusual $PM2.5$ states with large standard deviation and low means.

Generally, connected clusters of the same colors are more pronounced for $NO$ and $NO2$ as compared to $PM$.
This could be attributed to the long-range transport of the $PM$-particles by moving air. Patterns and spatiotemporal scales for $PM$ have been extensively discussed previously in \cite{Prospero:2001,Uno:2001,Bardouki:2003,Kallos:2006}. The transport depends on weather conditions and removes the memory to the site where the particles were originally produced. Thus the colors are more mixed in our plots.
As the weather patterns and the local meteorological features each contribute to the transport of $PM$-particles, the measuring site type has limited impact on the observed PDFs of $PM$. 

\subsection*{Fitting power-law tails for the data}

As exponential tails apparently do not fit the data well,
as illustrated by the deviations from the diagonal in Fig.2,
we now propose a different fitting function, motivated by many previous investigations in generalized versions of statistical mechanics \cite{tsallis2009introduction}. This is a fitting by a so-called
$q$-exponential, which asymptotically decays with a power-law exponent $-\frac{1}{q-1}$.
The normalized PDF is defined as follows:
\begin{equation}
f_{q,\lambda}(x) = (2 - q) \lambda [1 -\lambda (1 - q) x]^\frac{1}{1 - q} \text{ for } 1 -\lambda (1 - q) x \geq 0, x>0, 
\label{eq:qexp}
\end{equation}
where $q$ is the entropic index~\cite{tsallis2009introduction,hanel2011,jizba2019}, $\lambda$ is a positive width parameter and $x$, in our case, denotes the air pollutant concentration. Eq.~\eqref{eq:qexp} contains the exponential distribution as a special case, namely
for $q = 1$,
as the $q$-exponential function, defined as $e_{q}(x)= [1 + (1 - q) x]^\frac{1}{1 - q}$, converges to the exponential function in the limit $q\rightarrow 1$. For $q<1$, $f_{q,\lambda}(x)$  lives on a finite support and becomes exactly zero above a critical value x, since, by definition, $e_{q}(x) = 0$ for $1-\lambda(1-q)x<0$. In contrast, if $q > 1$, $1 -\lambda(1-q)x>0$, then eq.~(\ref{eq:qexp}) exhibits power-law asymptotic behavior. 

The occurrence of $q$-exponentials with $q>1$ in PDFs of
complex systems is very well-motivated by superstatistical models \cite{beck-cohen, BCS}.
In these types of models, one assumes a temporally fluctuating parameter $\lambda$ for local exponential distributions as given in eq.(1). These fluctuations of $\lambda$ take place
on a long time scale, much longer than local air pollution
concentration fluctuations.
The marginal distribution, obtained by integration over all possible values of $\lambda$, and
describing the long-term behaviour of the air pollution concentration dynamics, is then a  $q$-exponential, with 
\begin{equation}
    q= \frac{ \langle \lambda^2 \rangle}{\langle \lambda \rangle^2}.
\end{equation}
Here $\langle \cdots \rangle$ denotes the expectation
with respect to the PDF of $\lambda$, see \cite{beck-cohen} for more details. Strictly speaking, a $q$-exponential is only obtained exactly if $\lambda$ is $\Gamma$-distributed,
but the general idea of superstatistics is
that a parameter $q$ can be defined by eq.(4) for more general distributions different than the $\Gamma$ distribution as well. The concept that
wind changes and other effects (such as traffic fluctuations) can lead to a superstatistical dynamics for pollution concentrations was first worked out
in \cite{williams_superstatistical_2020}, where further details can be found, in that case for the special example of $NO$ and $NO2$ concentrations as measured in London. Our investigation here is much more general, as we include data of thousands of measuring stations, and also investigate $PM2.5$ and $PM10$ concentrations.

\begin{figure}[h]
    \includegraphics[width=13.5cm]{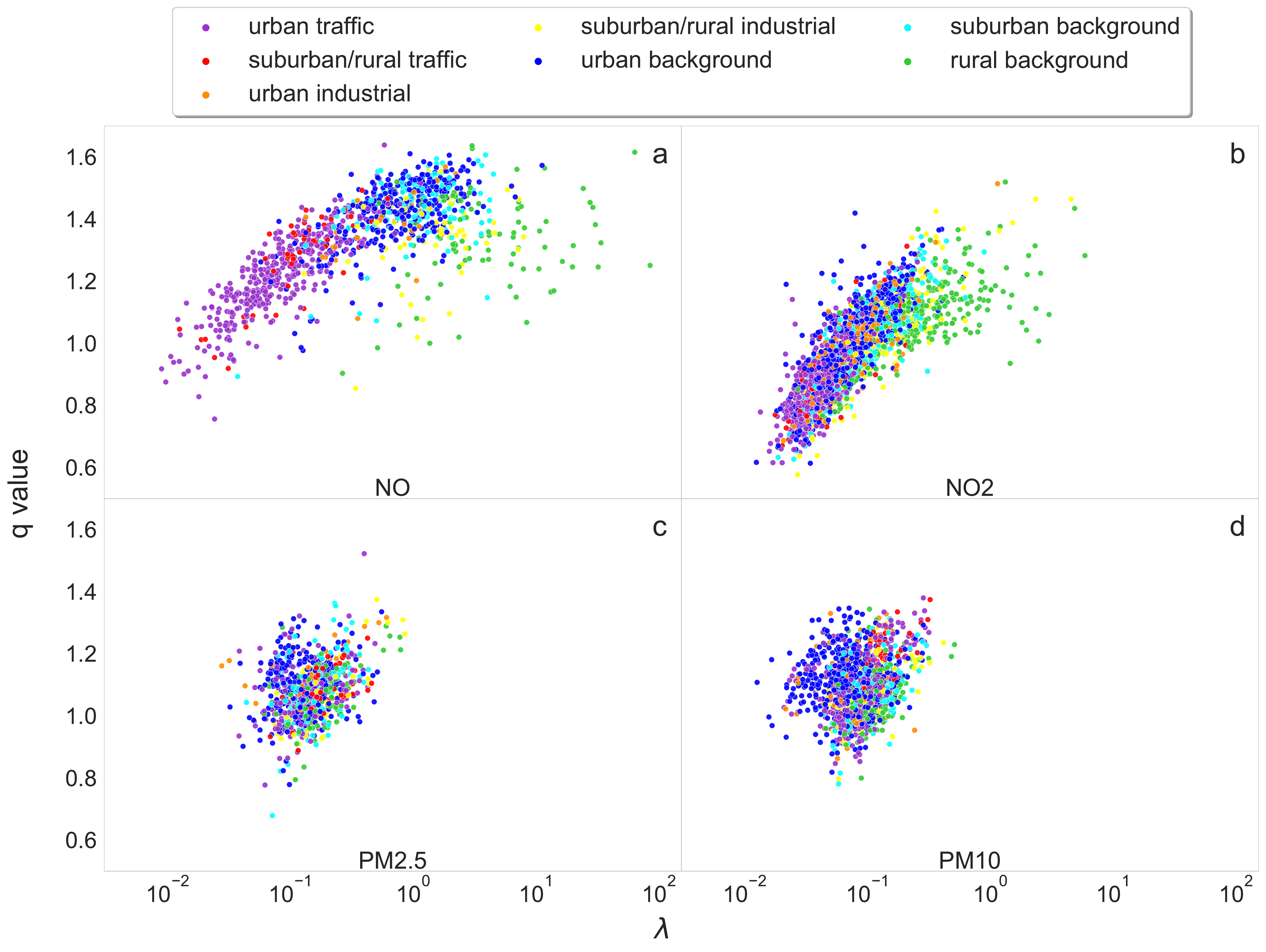}
    \caption{
    Best-fitting parameters of $q$-exponentials. We observe an increasing trend of $q$ versus $\log \lambda$ for $NO$ (a) and $NO2$ (b), whereas a more disk-shaped pattern is observed for $PM2.5$ (c) and $PM10$ (d). The environmental characterizations of the measuring stations are again encoded by colors. Again we observe  correlated patches of a single given color for $NO$ and $NO2$, where for $PM2.5$ and $PM10$ the patterns is more mixed.    \label{fig:q_vs_l}}
\end{figure}

For all our 3544 measuring stations we extract histograms
of the pollution concentration from the measured
time series,  and determine the  best-fitting parameters $q$ and $\lambda$ for the given data set. More details on the numerical procedure are described in the Method section. Our results are shown in Fig.~3.

A truly surprising result of our analysis is the fact that we observe an immensely large range of values of the parameter $\lambda$ for the best-fitting $q$-exponential as given in eq.(4) for the various measuring stations.
Note the logarithmic scale of the plots, the parameter $\lambda$ can take on values as small as $10^{-2}$ up to values as large as $10^2$, which spans four orders of magnitude. Typical $q$-values are in the range $0.8-1.4$, but there are subtle differences between the various substances, with $NO$ reaching large $q$-values such as 1.6, and $NO2$ reaching small $q$-values such as $0.6$ in the scattering plots. Also the shape of the scattering cloud of points is different for the different substances. For example, the typical range of $\lambda$ for $PM2.5$ and $PM10$  varies only by a factor 10, whereas for $NO$ and $NO2$ it varies by a factor $10^3$. The scattering plot data look more spherically symmetric for $PM2.5$ and $PM10$, as compared to $NO$ and $NO2$.  

Fig.~\ref{fig:q_vs_l} panels a and b indicate a roughly linear approximate relationship between $q$ and $\log \lambda$ for $NOx$, with most of the points corresponding to traffic clustered at the left, while urban/suburban background points are in the middle part, and rural background points are scattered widely at the right. The PDF decay rate increases as $\lambda$ increases from highly polluted urban traffic sites to less polluted rural background areas. For $PM2.5$ and $PM10$, there is a different weak uphill relationship between $q$ values and $\lambda$, as can be seen in Fig.\ref{fig:q_vs_l} panels c and d. The urban background points are reaching small $\lambda$ values such as $0.01$ for $PM10$, and suburban/rural traffic, suburban/rural industrial and rural background points cluster on the right hand side with large $\lambda$. 
The attained range of $\lambda$ values is smaller as compared to the case of NOx, and the shape of possible values $(q, \lambda)$ as displayed in the Figure is more spherical.
The colors appear to be more randomly mixed.

The stronger colour mixing for $PM2.5$ and $PM10$ can again be interpreted by the fact that by air movement transport the distributions cannot be uniquely identified with the original environmental types where the $PM$-particles were produced. The large scattering of parameters $(q, \lambda)$ shows that for a given substance at a given environmental type there is not just one possible distribution, but a large range of possible distributions. These distributions may also vary in time, according to the weather conditions. Still, our scattering plots allow us, in principle at least, to select {\em typical} parameter values for a given environmental type. The fact that there is a broad distribution of parameters is very much in line with the basic modelling assumption of superstatistics, in this case however applied to a spatial ensemble of different locations. There is
a strong heterogeneity in space, meaning different spatial
measuring locations have quite different PDFs. This spatial heterogeneity is an second effect, adding  to the temporal heterogeneity of local exponentials, which effectively leads to $q$-exponentials at individual locations as explained above.

\subsection*{Spatial distribution plots of $\lambda$-values}

\begin{figure}[h]
    \includegraphics[width=13.5cm]{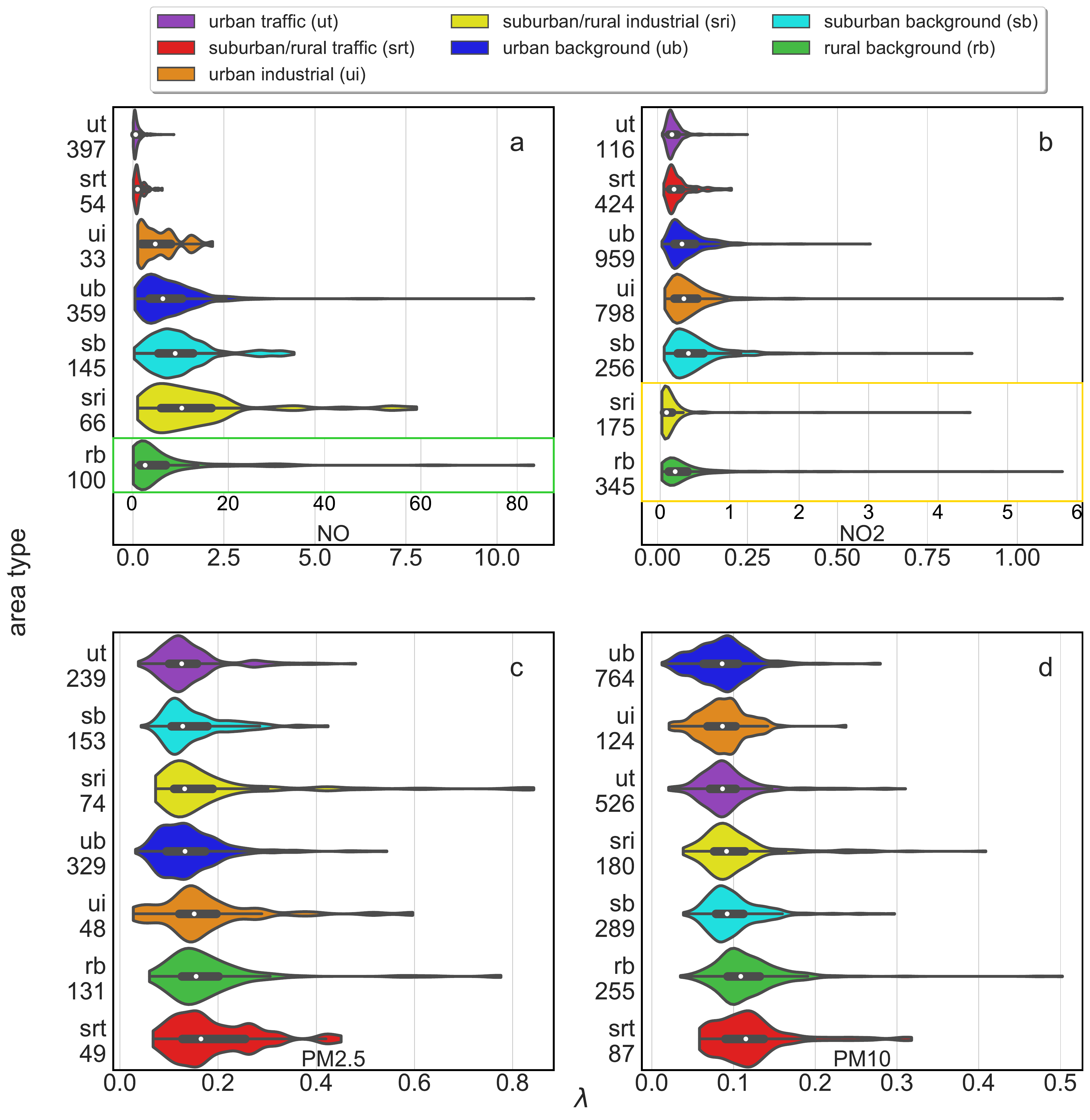} 
    \caption{
     The violin plots show the distribution of $\lambda$ for seven environment types as well as the median as a white dot, the interquartile range as a thick black bar, and the 95\% confidence interval as a thin black bar within the colored violin. The environment types were ranked by medians of $\lambda$ from lowest to highest. At each y-axis the number of sites evaluated in each category
     is reported. In the case of $NO$ (a), a rescaling has been applied to capture the different scale for rural background (green). Likewise, there is also a rescaling for suburban/rural industrial (yellow) and rural background for $NO2$.
    \label{fig:violin}}
\end{figure}

Finally, we are interested in the PDFs of $\lambda$
values for our fits of the various classified locations where the measurements are taken.
We compare the summary statistics (such as distribution, range and quartiles) of  $\lambda$ for the four air pollutants with the aid of so-called violin plots, see Fig.\ref{fig:violin}. Within these, we visualise the distribution of $\lambda$ using density curves, which corresponds to the approximate frequency of data. In addition to the density curves, an overlaid box plot is shown with the rectangle depicting the ends of the first (25\%) and third quartiles (75\%) and the central dot indicating the median (50\%).

An interesting result of the violin plots shown in Fig.~\ref{fig:violin} is that the $\lambda$ distributions extend to very large values, as indicated by the long  extensions to the right for area types such as urban background, suburban/rural industrial and rural background. Additionally, the probability distributions of $\lambda$ (represented by the "shapes" of the violins) exhibit nontrivial behaviour for some of the environmental site types. For example, for $NO$ urban industrial sites there is a rather unusual pattern with several local maxima and minima. A much more generic pattern, with just a single broad maximum, is observed for sites which are suburban/rural industrial, as well as for those with a rural background, and this structure is there for all 4 different types of pollutants. 

Another intriguing result is that the order of the median rankings (from low to high) for $NO$ and $NO2$ are almost the same, with the exception of a swap between urban industrial and urban background.
For $PM2.5$ and $PM10$ there are more swaps.
Fig.~\ref{fig:violin} panels c and d show that for $PM2.5$ and $PM10$ the typical values of $\lambda$ are smaller, below 0.9. A smaller $\lambda$ indicates a more heavily polluted site. Furthermore, we observe in the case of $PM$ only minor differences in the medians of $\lambda$ for different environment categories. The reason for this is that the type of environment has a direct effect on $NOx$ concentrations while they have only a minor effect on $PM$ since the particles travel
and lose the memory of their environmental category.
Nevertheless, suburban/rural traffic and rural background sites have the largest and second largest medians for $\lambda$, respectively. 


\section*{Discussion and Conclusion}

The use of $q$-exponentials for air pollution statistics has
been previously advocated by Williams et al.~\cite{williams_superstatistical_2020}, however that study was special since it was only looking at locations in Greater London, and the substances investigated were just $NO$
and $NO2$. In this paper, we have extended the statistical analysis
to a much larger database, taking into account data from 3544
measuring stations, and analysing the statistics of $PM2.5$ and $PM10$, in addition to $NO$ and $NO2$. Naturally,
for this vast amount of data novel methods needed to be developed for the fitting procedures, and novel graphical representations (scattering plots of parameter tuples) were 
used to illustrate the spatial heterogeneity of the results. Our main findings can be summarized as follows:

Firstly, we have clear evidence that
generically PDFs of pollution concentrations do not decay
in an exponential way. A much better fit is given
by $q$-exponentials, which asymtotically decay as a power-law if $q>1$, with exponent $-1/(q-1)$. Our analysis complements previous work, in which often
different distributions have been used, including log-normal, gamma and Weibull distributions. Overall, we find
that $q$-exponentials yield a better fit
of the tails. The $q$-exponential is also a very plausible physical model, since --in the spirit of superstatistics--
it simply arises from the agglomeration of many exponential
distributions that have temporal fluctuations of the effective
decay rate.
In our investigation we also tested the other candidate distributions mentioned above and found that they sometimes
yield a good fit of the low-concentration behaviour, close
to the maximum, but not of the tail behaviour, see our code for details.

Secondly, $q$ and $\lambda$, as obtained from the
optimum fittings of data from 3544 measuring stations,
exhibit interesting patterns in the $(q, \lambda)$ plane. The shape of these regions is characteristic for each of the 4 substances investigated, with big differences 
between $NO$, $NO2$ and the $PM$ statistics. 

Thirdly, environmental types, i.e. the surroundings
of the measuring station, play an important role.
We color-coded these different environments into
7 categories. For $NO$ and $NO2$ each category occupies
a typical sub-region in the $(q, \lambda)$ plane, whereas
for $PM2.5$ and $PM10$ the picture is more mixed.
This can be explained by the fact that there is transport
by moving air for the $PM$-particles, so that the memory
to the environment of the station where the actual measurements are done is lost, i.e. these particles may travel quite a long way and many of them are not produced locally.
The patterns and spatio-temporal scales for $PM$-dynamics have been extensively examined before in \cite{Prospero:2001,Uno:2001,Bardouki:2003,Kallos:2006}, the transport is dependent on the weather conditions and removes the memory of the original source site. 

As a next step, it would be desirable to examine correlations between $PM$ and $NOx$ concentrations and to include more environmental factors, including wind speed, wind direction and surface temperature. In light of these more detailed investigations, a statistical analysis could enable policymakers to produce more precise rules and thresholds 
for individual types of environmental conditions and meteorological conditions, taking into account fluctuations and extreme events. Our analysis could also be extended to other substances, such as sulfur oxide, carbon oxide and ozone, besides $NOx$ and $PM$. Taking all this into account could help policymakers to develop tailored guidelines to reduce health damage caused by air pollution. We stress again that the detailed description of the entire pollution spectrum, including the exact behaviour of the tails, seems critical here to better estimate the risks of very high pollution situations. 
Concluding, our analysis shows that there is strong heterogeneity in the data, and one needs to be careful because the PDFs vary
strongly from one location to another.


\section*{Methods}
\subsection*{Data processing}
We import European air pollution data from "Saqgetr"\cite{saqgetr}, which is an R package developed by Stuart Grange to analyse air quality data — or more generally atmospheric composition data. The package provides users with fast access to thousands of sites' data from air quality networks, which are supported by Ricardo Energy \& Environment. We import all 9698 sites' hourly data between 2017 and 2021 from the package and save them as individual .csv files. The raw data contain detailed information about the air pollution monitoring sites and their hourly measurements. From them, we select the following information:
\begin{enumerate}
    \item Name and site code. Each site has a unique site code for identification and simplification in coding. (E.g. "gb0050a" for the Rosia Road in Gibraltar)
 \item Longitude and latitude, which are used to show stations' locations on maps (such as Fig.~\ref{fig:map})
    \item $NO$, $NO2$, $PM2.5$ and $PM10$ pollutants' hourly concentration data
    \item Station and area types, which are combined for classifying sites
\end{enumerate}

The data sets retrieved with the Saqgetr package contain several problems, for which our solutions are:
\begin{enumerate}
    \item  We filter out sites with less than one year of $NO$, $NO2$, $PM2.5$ and $PM10$ pollution concentration data. We need a minimum number of 8760 data points ($365$ days $\times$  $24$ hours per day) for one pollutant for a meaningful statistical analysis and to avoid analysing a single season.
    \item Measurements below the detection limit, including some that are even below zero, are usually replaced by the detection limit divided by two. We filter out sites with more than $15\%$ of data below the detection limit. This is aligned with the recommendations of the US Environmental Protection Agency (US EPA)\cite{united2000guidance}, which suggests that substitution may be a viable approach when up to $15\%$ of the data cannot be detected. This steps is further justified as sites with extremely low pollution are not the main focus of this study.
    \item We remove sites whose data is repeating in a single measurement at least $15\%$ of the time. We assume that these sites either lack precision in measurement or contain too much corrupted data. 
\end{enumerate}

\subsection*{Fitting procedure}
\label{sec:Fitting_procedure}
\begin{figure}[h]
    \includegraphics[width=13.5cm]{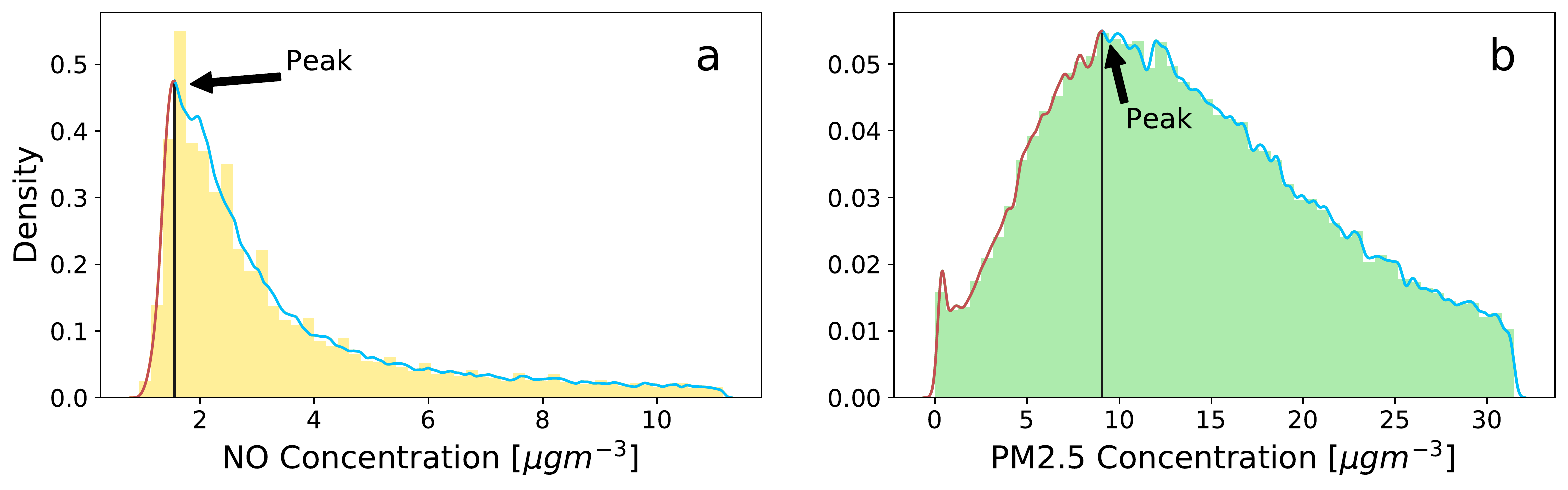} 
    \caption{
    Determination of the lower cutoff at peak density (vertical line). We display the empirical probability density functions (PDFs) of the $NO$ (a) and $PM2.5$ (b) concentrations with cut-off options at $1.5\times \mbox{mean}~\mu g/m^3$. The empirical smoothed PDF estimates the densities of the data before (red) and after the peak (turquoise).
    \label{fig:peak}}
\end{figure}

The filtered data sets were analysed in Python. First, we plot the probability density function (PDF) of each site's pollutant concentration level using a log scale. Then, we consider the distributions of the higher concentration tail by finding the maximum of the distribution and filtering out the smaller concentration distribution (left of the maximum of the distribution). One reason for doing so is that we are particularly interested in the high concentration tail behaviour. As an example, let us discuss the probability densities for low pollution concentrations for $NO$ in Amstetten, Austria (Fig.~\ref{fig:peak}(a)) and for $PM2.5$ in Riadok, Slovakia  (Fig.~\ref{fig:peak}(b)). We determine the highest point of the kernel line estimate and its corresponding concentration level (black vertical line). The underlying density distribution of the red line created an increasing slope which we cut off. We only analyse the underlying density distributions of the turquoise line and the higher concentration tail.

\par
We fit the so-obtained data with exponential and $q$-exponential distributions derived from MLE methods. C. Shalizi~\cite{shalizi2007maximum} and R. Gonzalez-Val~\cite{gonzalez} described methods for estimating the parameters of the $q$-exponential and log-normal distributions respectively using MLE. We carry out the fit, using the \emph{scipy.curve\_fit} and \emph{scipy.stats.rv\_continuous.fit} functions of the scipy module\cite{2020SciPy-NMeth}. The $q$-exponential distribution using MLE methods generally show robust fitting results. See code for further details and fits of non-$q$-exponential functions.

\subsection*{Data availability}
Data sets analyzed in the present study can be obtained via the saqgetr package from \url{https://github.com/skgrange/saqgetr}. They should be downloaded as csv files and named after their site codes. The code to generate the figures in the paper, as well as the implementation of the method for the data sets used in the paper, are available at \url{https://github.com/hurst0415/Spatial-heterogeneity-of-air-pollution-statistics}. 

\bibliographystyle{naturemag}
\bibliography{main}

\subsection*{Acknowledgements}
This project has received funding from the European Union's Horizon 2020 research and innovation programme under the Marie-Sklodowska-Curie Grant agreement No 840825. We also gratefully acknowledge funding from a 2022 QMUL Research England Policy Impact grant awarded to C.B. and funding from the Helmholtz Association under grant no. VH-NG-1727 to B.S.

\subsection*{Author contributions}
H.H., B.S. and C.B. conceived the project, H.H. and B.S. processed the data,  performed the data analysis and produced the plots. All authors interpreted the results, wrote and reviewed the manuscript.
Overall B.S. and C.B. contributed equally.

\subsection*{Competing interests}
The authors declare no competing interests.

\end{document}